\documentclass[referee]{aa} % for a referee version

\usepackage{graphicx}
\usepackage{txfonts}

\begin{document}

\title{Comparison of counterstreaming suprathermal electron signatures of ICMEs with and without magnetic cloud: are all ICMEs flux ropes?}
\author{Jiemin Wang, Yan Zhao, Hengqiang Feng, Qiang Liu, Zhanjun Tian, Hongbo Li, Ake Zhao, Guoqing Zhao}
\offprints{Hengqiang Feng}
\institute{Institute of space physics, Luoyang Normal University, Luoyang 471934, China\\
              \email{fenghq9921@163.com}
            }

\abstract
% context heading
{Magnetic clouds (MCs), as large-scale interplanetary magnetic flux ropes, are usually still connected to the sun at both ends near 1 AU. Many researchers believe that all non-MC interplanetary coronal mass ejections (ICMEs) also have magnetic flux rope structures, which are inconspicuous because the observing spacecraft crosses the flanks of the rope	structures. If so, the field lines of non-MC ICMEs should also be usually connected to the Sun on both ends.}
% aims heading
{Then we want to know whether the field lines of most non-MC ICMEs are still connected to the sun at both ends or not.}
% methods heading
{This study examined the counterstreaming suprathermal electron (CSE) signatures of 266 ICMEs observed by the \emph{Advanced Composition Explorer} (\emph{ACE}) spacecraft from 1998 to 2008 and compared the CSE signatures of MCs and non-MC ICMEs.}
% results heading
{Results show that only 10 of the 101 MC events ($9.9\%$ ) and 75 of the 171 non-MC events ($43.9\%$) have no CSEs. Moreover, 21 of the non-MC ICMEs have high CSE percentages (more than $70\%$) and show relatively stable magnetic field components with slight rotations, which are in line with the expectations that spacecraft passes through the flank of magnetic flux ropes. So the 21 events may be magnetic flux ropes but the \emph{ACE} spacecraft passes through their flanks of magnetic flux ropes.}
% conclusions heading
{Considering that most other non-MC events have disordered magnetic fields, we suggest that some non-MC ICMEs inherently have disordered magnetic fields, namely have no magnetic flux rope structures.}


\authorrunning{Wang, et al.}
\titlerunning{Ccounterstreaming suprathermal electron signatures of ICMEs}

\maketitle

\section{Introduction}

Coronal mass ejections (CMEs) are intense solar explosive eruptions during which large amounts of plasma and magnetic fields from the solar atmosphere are ejected to the interplanetary space. The interplanetary manifestations of CMEs (ICMEs; Kilpua et al. 2017) can be measured by a spacecraft at about 1 AU and exhibit the following characteristics: increase in total magnetic magnitude (Cane \& Richardson, 2003), helium abundance (Hirshberg et al. 1972; Zwickl et al. 1983; Richardson \& Cane 2004.), average iron ionization (Lepri et al. 2001; Lepri \& Zurbuchen 2004), and $O^{7+}$ abundance (Richardson \& Cane 2004, Wang \& Feng 2016); decrease in proton temperatures and proton densities (Gosling et al. 2001; Zhang et al. 2013); counterstreaming suprathermal electron (CSE) strahls and declining speed (Zwickl et al. 1983; Gosling et al. 1987; Shodhan et al. 2000; Burlaga et al. 2001).
A subset of ICMEs was defined as magnetic cloud (MC) by Burlaga et al. (1981) empirically using the following properties: (1) the magnetic field strength is higher than average, (2) a smooth change in field direction
as observed by a spacecraft passing through the cloud, and (3) low proton temperature compared to the ambient proton temperature. MCs usually have magnetic flux rope structures, and they are the main source of major geomagnetic storms (Burlaga et al. 1981; Webb et al. 2000; Huttunen et al., 2002; Zhang et al., 2007). Observations at 1 AU show that $30\%$-$40\%$ of ICMEs are MCs, and this percentage depends on the solar cycle (Richardson \& Cane, 2004). However, CMEs are usually assumed to have magnetic flux rope structures near the sun because of their helical shapes (Canfield et al. 1999; Liu et al. 2010; Rust \& Kumar 1996). Thus, do non-MC ICMEs also have flux rope structures? The journal of \emph{Solar Physics} once made a special issue to address this question (Gopalswamy et al. 2013a). A comparative study of 23 MCs and 31 non-MC ICMEs was completed, and the source regions of the 54 ICMEs were located within $\pm$15$^{o}$ longitude from the disk center. Yashiro et al. (2013) found that the structures of the post-eruption arcades of MCs and non-MC ICMEs during launch have no significant difference. Gopalswamy et al. (2013b) observed that MCs and non-MC ICMEs have significant enhancement in \emph{Fe} and \emph{O} charge states, and \emph{Fe} and \emph{O} charge-state measurements are positively correlated with flare properties, including flare size and soft X-ray flare intensity. Their observations suggest that these CMEs have similar explosive environment and flux rope structures near the sun. Furthermore, some studies indicate that CMEs associated with MCs tend to propagate along the sun-Earth line, whereas non-MC events are deflected away from the sun-Earth line (Kim et al. 2013; M\"{a}kel\"{a} et al. 2013; Zhang et al. 2013). Therefore, many researchers believe that all ICMEs have magnetic flux rope structures and that the non-MC events are due to observational limitations, that is, observing spacecraft crosses the flanks of the ropes and thus the ICMEs appears as non-MCs. This has been shown  by some multi-satellite observed ICMEs, namely spacecraft farther from the axis detects less clear flux rope signatures than centrally crossing spacecraft for the same event(Cane et al., 1997; Kilpua et al., 2011).

Moreover, some researchers believe that some ICMEs may have lost their flux rope structure due to interactions in interplanetary space (e.g., Gopalswamy et al., 2001; Liu et al., 2014; Manchester et al., 2017). In particular, sometimes multiple ICMEs can merge to form a complex ejecta/ICMEs where their individual characteristics are not identifiable anymore (Burlaga et al. 2002). According to Rodkin et al. (2018), about $48\%$ of observed ICMEs are associated with two or more sources; in some cases, complex ICMEs can be associated with multiple CMEs from the same active region (Burlaga et al. 2002). The interaction processes include compression and magnetic reconnection; however, the compression process does not change the overall topology of rope structures (Riley \& Crooker, 2004, Zhang et al. 2013). Therefore, the rope structures of CMEs are destroyed mainly through magnetic reconnection processes. A CME that was associated with complex (non-MC) ejecta may inherently had complex magnetic field structure at the Sun, as shown in the simulation results of Lynch et al. (2008). Richardson \& Cane (2004) favored the view that the reconnection of multiple loop systems may result in CMEs with several complicated magnetic field structures as solar activity increases. This explanation is appropriate for the observation that a fraction of MCs vary with the phase of the solar cycle, that is, about $15\%$ at solar maximum but almost $100\%$ at solar minimum. However, CME source regions cluster to low latitudes at solar minimum and high latitudes at solar maximum (Hundhausen et al. 1984), thus affecting whether CMEs are expected to be crossed more centrally or at flanks. Recently, Awasthi et al. (2018) reported a non-MC ICME that supports this view. Its pre-eruptive structure consists of multiple-braided flux ropes with different degrees of coherency, and the individual flux-rope branches manifest reconnection with each other. Awasthi et al. (2018) excluded that the complex ICME can result from the merging of successive CMEs and the spacecraft they used only made a glancing encounter with Earth-directed CME. Awasthi et al. (2018) considered that a non-MC event inherently disrupts magnetic fields.

Suprathermal electron strahls in the solar wind come from the Sun and are focused along magnetic field lines (Feldman et al. 1975; Rosenbauer et al. 1977; Pagel et al. 2005). Therefore, observations of CSE strahls within MCs can indicate that the flux rope structures are still connected with the sun¡¯s magnetic field lines on both ends (Gosling et al. 1995; Larson et al. 1997; Shodhan et al. 2000; Feng et al. 2015, 2019). Therefore, observations of CSE strahls within MCs can indicate that the flux rope structures are still connected with the sun¡¯s magnetic field lines on both ends (Gosling et al. 1995; Larson et al. 1997; Shodhan et al. 2000; Feng et al. 2015, 2019). CSEs can also be produced by other mechanisms, e.g., connection to the Earth¡¯s bowshock (Feldman et al., 1982; Stansberry et al., 1988), interplanetary shocks or CIRs (Gosling et al., 1993; Steinberg et al., 2005; Lavraud et al., 2010) and distribution function depletions near the 90$^{o}$ pitch angle (Gosling et al., 2001, 2002; Skoug et al., 2006). Among these mechanisms, the depletion CSEs are often observed on closed or open field lines within ICMEs, but the depletion CSEs are centered on and roughly symmetric about 90$^{o}$ pitch angle (Gosling et al. 2002), and can be distinguished from CSE strahls. Shodhan et al. (2000) examined the CSE strahl signatures of 52 MCs detected by using a spacecraft near 1 AU and determined that approximately $87.5\%$ of MCs exhibit CSE signatures, revealing that most MCs are still attached to the sun on both ends at 1 AU. Gosling et al. (1990, 1995) have proposed explanations for how flux ropes arise in terms of 3-dimensional reconnection close to the Sun. They illustrate the original flux rope reconnect to form a helical field line connected to the Sun at both ends. The closed field lines of flux ropes can gradually open and occasionally disconnect from the corona when closed flux ropes expand from the Sun into the interplanetary space (Gosling et al. 1995). The proposal of Gosling et al. (1995) was confirmed by statistical results of Shodhan et al. (2000), namely most MCs exhibit CSE strahls in parts of their durations, only 6 of the 52 MCs have no CSE. Then we want to know whether non-MC ICMEs have the same CSE strahl signatures. In this study, CSE signatures from \emph{Advanced Composition Explorer} (\emph{ACE}) during 1998-2008 are compared between ICMEs with and without MCs, and aim to discuss whether the CSE signatures are related to the flux-rope structures.

\section{ Data}

In this study, the 272 eV suprathermal electron pitch-angle distributions (PADs) measured by ACE are used. The electron PADs are obtained from the \emph{Solar Wind Electron Proton Alpha Monitor} (\emph{SWEPAM}) with angular resolution and time resolution at 9 degrees and 64 second (McComas et al., 1998). This study examined 16-s average magnetic field, 64-s average plasma, 1-h average $O^{7+}$/$O^{6+}$ ratio, and mean Fe charge state $\langle$ Fe $\rangle$ data from 1998 to 2008 measured by \emph{ACE} and identified 272 ICMEs in total. The ICMEs were identified the following process: (1) We take the events in previous ICMEs lists of Jian et al. (2006), Chi et al. (2016), Richardson \& Cane (http://www.srl.caltech.edu/ACE/ASC/DATA/level3/icmetable2.htm\#(g)) as candidate ICMEs. (2) Some lists report also short-duration ($<$10 h) structures as ICMEs. As the origin of these smaller scale ICMEs and flux ropes are still debated (Feng et al. 2007; Rouillard et al. 2011; Janvier et al. 2014; Feng \& Wang 2015; Wang et al. 2019), we have excluded them from this study. (3) The high Fe charge states ($\langle$ Fe $\rangle\geq$12), abnormally high $O^{7+}$/$O^{6+}$ ratio ($\geq$1) are the result of flare-related heating in the corona (Lepri \& Zurbuchen 2004; Reinard, 2005), so they are independently reliable ICME indicators (Feng \& Wang 2015). If the candidate ICMEs have high Fe charge states or$/$and abnormally high $O^{7+}$/$O^{6+}$ ratio, they are identified as ICMEs.  (4) If the candidate ICMEs have no high Fe charge states and abnormally high $O^{7+}$/$O^{6+}$ ratio, we will examine the following 5 characteristics: declining speed (apparent expansion), increase in total magnetic magnitude, helium abundance (He/P$>$0.06) (Richardson \& Cane 2004), decrease in proton temperatures and proton densities. If the candidate ICMEs have 3 or more above characteristics, they are identified as ICMEs. Given that magnetic flux ropes are special field topologies characterized by bundles of helical magnetic-field lines collectively spiraling around a common axis, the essential observational properties of magnetic flux ropes should enhance magnetic field strength and field smooth rotations (Feng et al. 2008; Feng et al. 2010), namely, measured enhanced magnetic field strength, the center-enhanced magnetic components and bipolar curve magnetic components. Therefore, if an ICME have the enhanced magnetic field strength, both center-enhanced and bipolar field components, it was identified as a MC. Among the 272 ICMEs, 101 (37.1\%) events are identified as MCs. All 272 ICMEs are listed in Table 1, the second and third columns show the start and end times, the fourth column gives the duration of the ICMEs, and the fifth column provides the types of ICMEs (MC or non-MC).

\section{Observations and Results}

This study analyzed the 272 eV suprathermal electron PADs to determine the CSE signatures within ICMEs. The CSE events were confirmed through their suprathermal electron PADs that show significantly higher phase space densities near 0$^{o}$ and 180$^{o}$ pitch angle directions than in the 90$^{o}$ pitch angle direction (Lavraud et al. 2010). If the ratios of phase space densities near 0$^{o}$ and 180$^{o}$ to the phase space densities at 90$^{o}$ in an interval are greater than 3, we will take the interval as possible CSE interval. Then we the remove depletion CSEs on open field lines by eyes using their symmetries. Last, the phase space densities at 90$^{o}$ can obviously strengthen by scattering of strahl electrons. For some intervals within ICMEs, the ratios of phase space densities near 0$^{o}$ and 180$^{o}$ to the phase space densities at 90$^{o}$ are less than 3, but the CSE strahls still can be recognized by eyes, we will also take these intervals as CSE strahls. CSE boundaries are not always recognizable, and thus the identification of CSE intervals within ICMEs is somewhat subjective (Shodhan et al. 2000). We used the ICME in October 21-22, 1999 as an example to illustrate the process of identifying a CSE interval.

\begin{figure*}
\sidecaption
\resizebox{\hsize}{16cm}{\includegraphics[width=10cm]{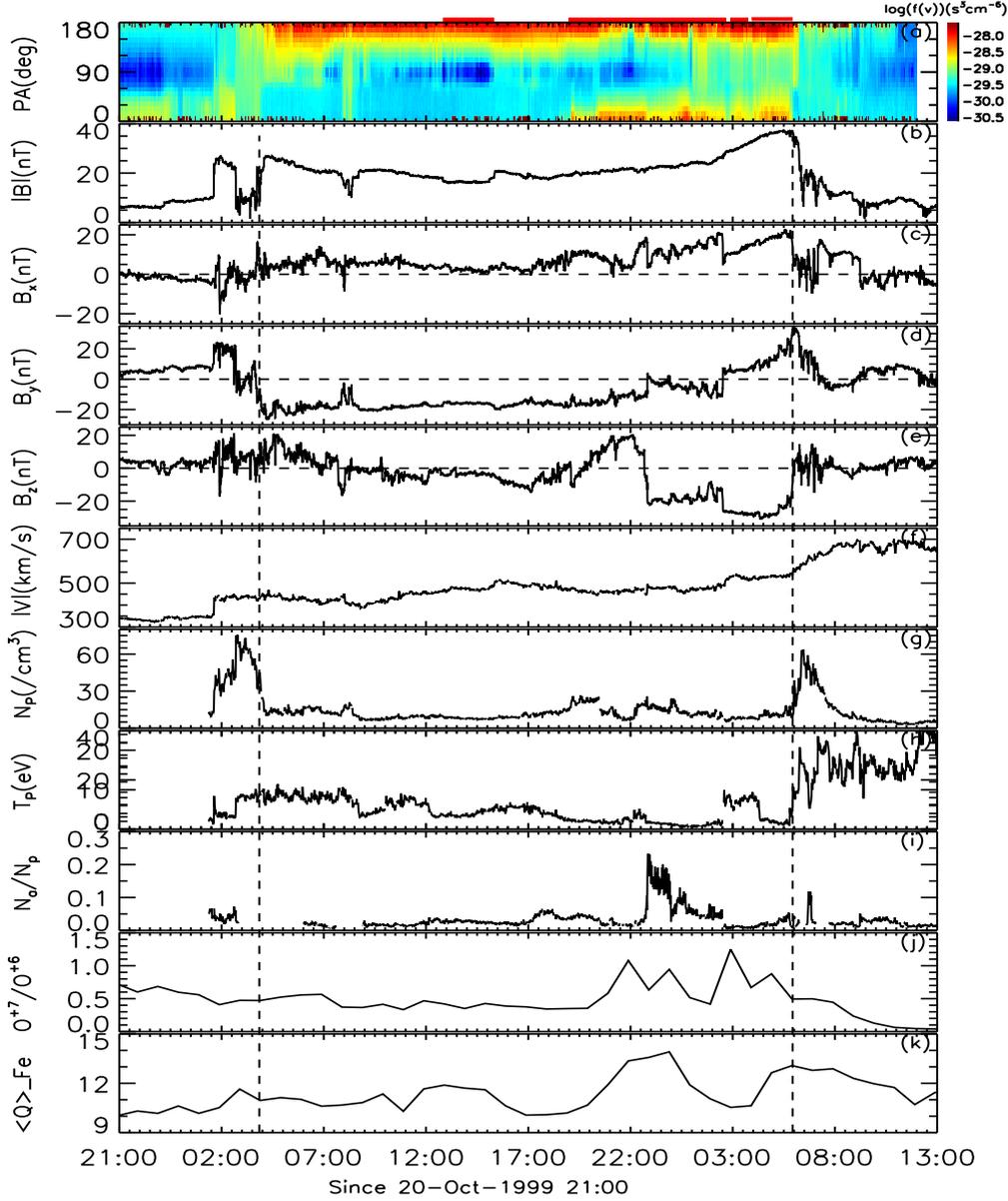}}
\caption{ Suprathermal electron pitch angle distributions of 272 eV, magnetic field and plasma data measured by \emph{ACE} during the October 21¨C22, 1999 ICME passage. The two vertical dashed lines denote the boundaries of the ICME.}
\end{figure*}

Figure 1 shows the suprathermal electron PADs, magnetic and plasma data in October 21-22, 1999 ICME passage. The two vertical-dashed lines represent the front and rear boundaries of ICME. This ICME exhibits high Fe charge states ($\langle$ Fe $\rangle\geq$12), abnormally high $O^{7+}$/$O^{6+}$ ratio ($\geq$1) and high He/P($>$0.06) in the rear part. In addition, the event also shows the ICME characteristics of enhanced magnetic magnitude, decreased proton temperatures and proton densities. The 272 eV suprathermal electron PADs showed that (1) the enhanced phase-space densities near 180$^{o}$ directions occurred throughout the duration of the ICME; (2) the 0$^{o}$ suprathermal electron strahls mainly present in the rear half part. The red bars above the suprathermal electron PADs marks the intervals where the ratios of phase space densities near 0$^{o}$ and 180$^{o}$ to the phase space densities at 90$^{o}$ are greater than 3. It is easy to find that strong depletions occurred within the first bar, and the depletion CSE interval was removed when we estimated the percentages of CSE. One can also find that the bars in the rear half part have two short intervals, which are caused by scattering of strahl electrons. So we take the two short intervals as CSE strahls, and the percentage of CSE strahls is about 35\%.

\begin{figure*}
\sidecaption
\resizebox{\hsize}{10cm}{\includegraphics[width=16cm]{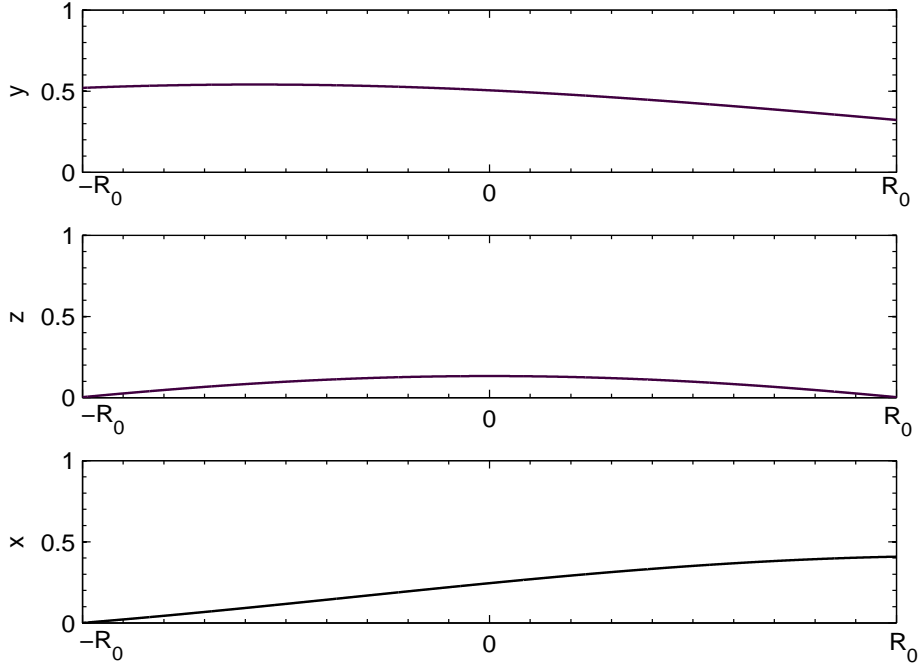}}
\caption{ Variation of normalized magnetic field components for the case of $d_{0}=0.9R_{0}$, where $R_{0}$ is radius of the flux rope, $d_{0}$ is the distance between the spacecraft trajectory and the rope axis.}
\end{figure*}

Furthermore, this study examined all the suprathermal electron PADs of the 272 ICMEs and estimated the percentages of CSEs. The results are listed in the sixth column of Table 1. Only 10 (9.9\%) of the 101 MCs have no CSE, and the ratio is consistent with the results of Shodhan et al. (2000), who examined the CSE signatures of 52 MCs and concluded that 6 MCs have no CSE. As for the other 91 MCs, the percentages of CSE vary from 6\% to 100\%, with a mean value of 57\%. Meanwhile, 75 of the 171 non-MC ICMEs (43.9\%) do not contain CSEs. The fraction of ICMEs without CSE is thus clearly larger in non-MC ICMEs than in MCs. The extensive difference may imply that the two groups have different magnetic structures. For the other 96 non-MC events, their mean CSE percentage is 45\%, which is smaller than 57\%. Among the 96 non-MC ICMEs, 21 have a percentage greater than 70\%.  An examination of the magnetic field component curves of these 21 ICMEs reveal that fields of most events have neither apparent smooth rotations nor disorder, but their magnetic field component curves have only slight rotations, and they are relatively stable or even close to the line. As we all know that most MCs can be described with constant a force-free field configuration, e.g., Lundquist solution (Lepping et al. 1990; Feng et al. 2006 ). According to the Lundquist solution, the shapes of measured field component curves depend on the distance $d_{0}$ between the spacecraft trajectory and the rope axis (See Figure 2, 3, 4 of Feng et al. 2006). The spacecraft will measure center-enhanced and bipolar curves if the spacecraft trajectory are close the rope axis, or else, the bipolar curves will disappear. Figure 2 shows the variations of the magnetic field components for the case of $d_{0}=0.9R_{0}$, where $R_{0}$ is radius of the flux rope; \emph{z} is the axial direction, \emph{x} is the radial direction pointing to the satellite contact position, the direction \emph{y} is obtained from the cross product of \emph{z} and \emph{x}. In Figure 2, all the three magnetic field components are relatively stable and exhibit only slight rotations. These trends indicate that the spacecraft that passes through the flank of a magnetic flux ropes will measured relatively stable magnetic field variations. So the 21 highest percentage (more than 70\%) events may have flux rope structures, but the spacecraft crosses the flanks of the ropes and thus the flux rope signatures didn't appear. For example, Figure 3 shows that the ICME in February 18-19, 2005 has a CSE percentage of 100\%. Based on Figure 3, one can find that all three magnetic field component curves remained stable with little fluctuation; in particular, the \emph{z} component curve is around zero. Meanwhile, the magnetic field magnitude remains at a low value. This event does not satisfy the requirements of MCs (flux ropes). Kim et al. (2013) investigated the propagation characteristics of this ICME and used the direction parameter (D) to quantify the asymmetry of CME shapes in coronagraph images and the degree of deviation from the sun-Earth line. The value of D is between 0 and 1, and a small D indicates a large deviation from the sun-Earth line. For this ICME, the D value is only 0.13, indicating that its propagation direction is essentially deviated from the sun-Earth line. Thus, Kim et al. (2013) considered that this CME inherently has a magnetic flux rope structure even though the ACE spacecraft crossed the flank of the rope and did not measure the essential properties of the magnetic flux rope. For this ICME, we have the same viewpoint with Kim et al. (2013). Figure 2 has demonstrated that spacecraft passes through the flank of magnetic flux ropes will measure relatively stable magnetic field variations. The MC ¡°legs,¡± which magnetically connect the flux rope to the Sun, are not recognizable as MCs and thus are unlikely to contain twisted flux rope fields. Spacecraft encounters with these non-flux rope legs may provide an explanation for the frequent observation of non-MC ICMEs. It is also possible that the spacecraft passes through the legs of the flux rope where the flux rope like rotation is not observed but other flux rope signatures are present (Owens et al., 2016). So we concluded that most of the 21 ICMEs with large CSE percentages may inherently have flux rope structures, but the structures did not appear as MCs because of geometric selection effects. Finally, we can draw such a conclusion that only a few MCs do not exhibit CSE signatures, but about half of the non-MC ICMEs do not exhibit CSE signatures.
\begin{figure*}
\sidecaption
\resizebox{\hsize}{16cm}{\includegraphics[width=10cm]{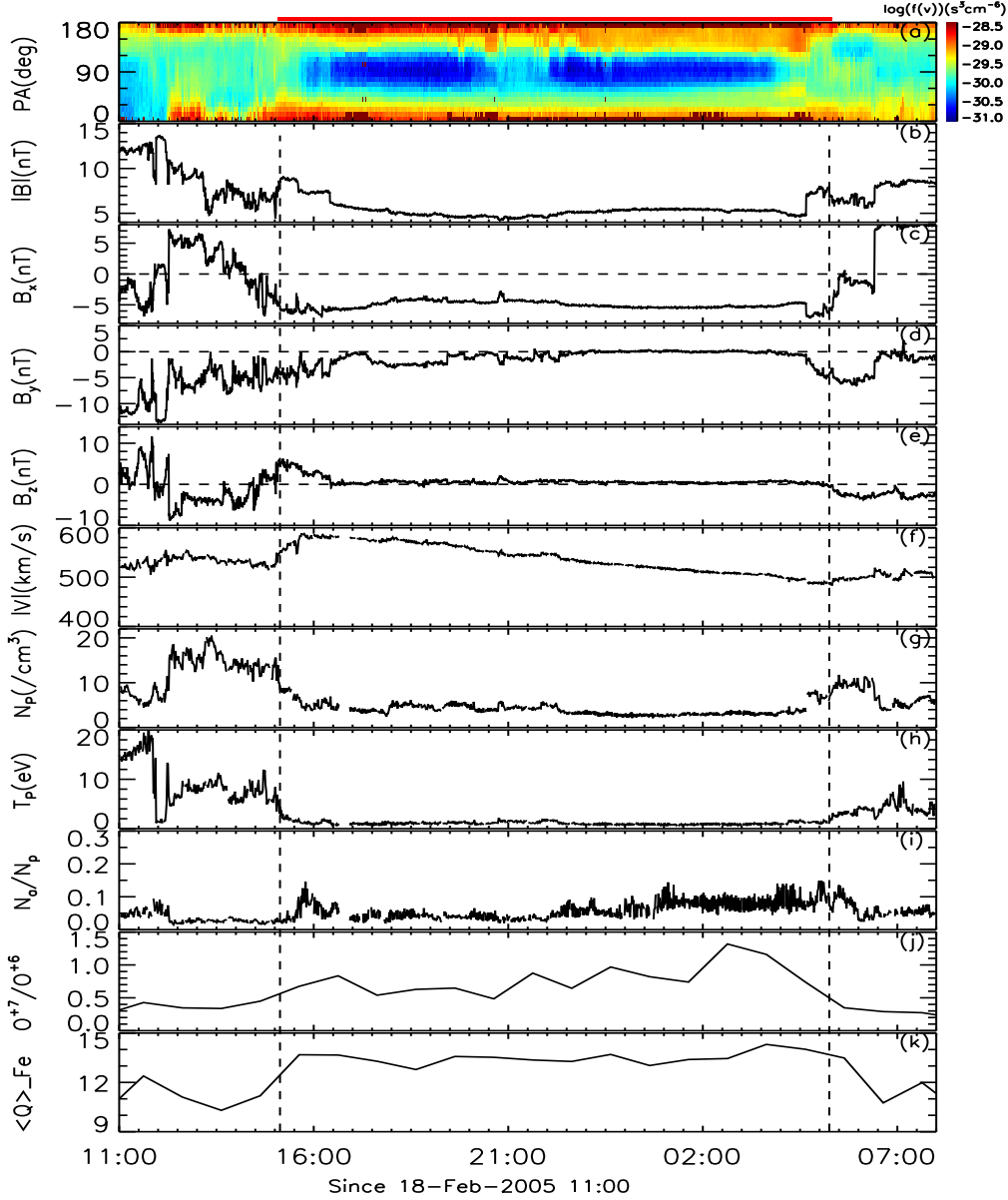}}
\caption{ Suprathermal electron pitch angle distributions of 272 eV, magnetic field and plasma data measured by \emph{ACE} during the February 18¨C19, 2005 ICME passage.}
\end{figure*}

\section{Discussion and summary}

In this study, we have examined the CSE signatures of the 272 ICMEs detected by ACE from 1998 to 2008 and compared them between ICMEs with and without MC (flux rope) structure. Our study confirms the earlier results e.g. by Shodan et al. (2000) that the overwhelming majority of MCs are still connected to the solar magnetic field on both ends for at least parts of their magnetic field lines near 1 AU. The clear majority of MCs exhibiting CSE suggests that the non-MC population featuring CSE could also have flux rope structure. We find that the MC category had 101 events, and only 10 (9.9\%) events had no CSEs, the non-MC category had 171 events, but 75 (43.9\%) event had no CSEs. The fraction of non-MC events without CSE is distinctly larger than that for the MC events. The highest percentage (more than 70\%) events in the non-MC category usually have stable magnetic field components accompanied with slight rotations based on our visual evaluation. These observations are in line with the expectations that spacecraft passes through the flank of magnetic flux ropes. The results of this study imply that some non-MC events indeed have magnetic flux rope structures. In most non-MC events, magnetic fields are disordered, and most field lines are not connected to the sun at both ends. Like reported CME by Awasthi et al. (2018), these non-MC events may inherently have disordered magnetic fields, which results from the interactions among multiple-braided flux ropes with different degrees of coherency on the sun. In summary, this study provides information that is helpful in checking whether all ICMEs have flux rope structures. The observations in this study do not support the idea that all CMEs that arrive on Earth have flux ropes but support that some non-MC events indeed have a magnetic flux rope structures. However, we can not exclude the possibility that all CMEs that arrive to Earth would have had a flux rope structure when they were launched from the Sun. The association of the CSE signatures and the CME structures is an interesting problem and can be investigated in future.

\begin{acknowledgements}

The authors acknowledge supports from NSFC under grant Nos. 41804162, 41674170, 41974197.  The authors thank NASA/GSFC for the use of data from ACE, these data can obtain freely from the Coordinated Data Analysis Web (http://cdaweb.gsfc.nasa.gov/cdaweb/istp$\_$public/).
\end{acknowledgements}

\clearpage

\begin{table}
\caption{The percent counterstreaming suprathermal electron in ICMEs.}
\begin{flushleft}
\begin{tabular}{lcccccc}
\hline
\multicolumn{1}{c}{NO.}&Start$^{a}$&End$^{b}$&Duration$^{c}$&Type$^{d}$&Percent$^{e}$\\
\hline

001&        1998/01/07 02:50&         1998/01/08 13:08&          34.3&       MC     &         56 \\
002&        1998/01/21 05:40&         1998/01/22 13:22&          31.7&       non-MC &         6  \\
003&        1998/01/29 19:56&         1998/01/30 23:08&          27.2&       non-MC &         23 \\
004&        1998/02/02 13:10&         1998/02/04 02:15&          37.1&       non-MC &         0  \\
005&        1998/02/04 04:15&         1998/02/05 15:10&          34.9&       MC     &         20 \\
006&        1998/02/17 10:00&         1998/02/17 21:00&          11.0&       MC     &         88 \\
007&        1998/02/19 00:20&         1998/02/20 00:10&          23.8&       non-MC &         0  \\
008&        1998/03/04 13:00&         1998/03/06 05:40&          40.7&       MC     &         0  \\
009&        1998/03/25 12:00&         1998/03/26 09:50&          21.8&       MC     &         93 \\
010&        1998/04/01 01:45&         1998/04/03 01:35&          47.8&       non-MC &         8  \\
011&        1998/04/12 01:20&         1998/04/13 17:50&          40.5&       non-MC &         0  \\
012&        1998/05/02 11:47&         1998/05/04 02:30&          38.7&       MC     &         53 \\
013&        1998/05/05 07:05&         1998/05/06 23:20&          40.3&       non-MC &         32 \\
014&        1998/06/14 05:10&         1998/06/15 07:05&          25.9&       MC     &         56 \\
015&        1998/06/24 12:00&         1998/06/25 23:15&          35.3&       MC     &         51 \\
016&        1998/06/26 07:10&         1998/06/26 18:58&          11.8&       non-MC &         94 \\
017&        1998/07/06 05:40&         1998/07/09 07:12&          73.5&       non-MC &         36 \\
018&        1998/07/11 13:54&         1998/07/12 21:40&          31.8&       non-MC &         0  \\
019&        1998/08/01 14:38&         1998/08/03 09:38&          43.0&       non-MC &         0  \\
020&        1999/08/04 03:37&         1999/08/05 12:30&          33.4&       non-MC &         33 \\
021&        1998/08/10 07:15&         1998/08/11 21:52&          38.6&       non-MC &         08 \\
022&        1998/08/11 23:50&         1998/08/13 13:12&          37.4&       non-MC &         53 \\
023&        1998/08/20 07:58&         1998/08/21 19:22&          35.4&       MC     &         27 \\
024&        1998/08/26 21:30&         1998/08/28 00:15&          26.8&       non-MC &         57 \\
025&        1998/09/23 04:02&         1998/09/23 17:45&          13.7&       MC     &         38 \\
026&        1998/09/25 06:04&         1998/09/26 16:05&          34.0&       MC     &         100\\
027&        1998/10/19 03:53&         1998/10/20 07:45&          27.9&       MC     &         63 \\
028&        1998/10/23 15:45&         1998/10/24 16:10&          24.6&       non-MC &         90 \\
029&        1998/11/08 04:20&         1998/11/09 02:55&          22.6&       MC     &         31 \\
030&        1998/11/09 02:55&         1998/11/10 06:42&          27.8&       MC     &         46 \\
031&        1998/11/13 00:58&         1998/11/14 13:50&          36.9&       MC     &         34 \\
032&        1998/11/30 08:56&         1998/12/01 02:40&          17.7&       non-MC &         20 \\
033&        1998/12/29 18:35&         1998/12/31 01:00&          30.4&       non-MC &         46 \\
034&        1999/01/04 04:05&         1999/01/04 22:19&          18.2&       non-MC &         41 \\
035&        1999/01/13 09:56&         1999/01/14 15:31&          29.6&       non-MC &         0  \\
036&        1999/01/23 04:50&         1999/01/23 17:38&          12.8&       non-MC &         30 \\
037&        1999/02/13 18:45&         1999/02/14 15:30&          20.8&       non-MC &         0  \\
038&        1999/02/16 14:18&         1999/02/17 08:54&          18.6&       non-MC &         0  \\
039&        1999/02/17 11:40&         1999/02/18 09:42&          22.0&       non-MC &         0  \\
040&        1999/02/18 09:42&         1999/02/19 10:41&          25.0&       MC     &         90 \\
041&        1999/02/19 22:37&         1999/02/20 18:05&          19.5&       non-MC &         0  \\
042&        1999/03/10 17:38&         1999/03/12 02:02&          32.4&       non-MC &         33 \\
043&        1999/03/19 10:34&         1999/03/20 16:20&          29.8&       non-MC &         41 \\
044&        1999/04/16 17:56&         1999/04/17 18:57&          25.0&       MC     &         63 \\
045&        1999/04/21 04:20&         1999/04/22 13:42&          33.4&       MC     &         79 \\
046&        1999/05/16 08:45&         1999/05/18 00:03&          39.3&       non-MC &         0  \\
047&        1999/06/03 00:17&         1999/06/03 21:43&          21.4&       non-MC &         0  \\
048&        1999/06/28 04:08&         1999/06/29 03:02&          22.9&       non-MC &         0  \\
049&        1999/07/03 06:37&         1999/07/05 13:38&          55.0&       non-MC &         0  \\
050&        1999/07/06 21:25&         1999/07/07 16:53&          19.5&       non-MC &         52 \\
051&        1999/07/08 03:58&         1999/07/08 22:08&          18.2&       non-MC &         0  \\
052&        1999/07/27 17:45&         1999/07/29 09:43&          40.0&       non-MC &         91 \\
053&        1999/07/30 11:40&         1999/07/31 08:20&          20.7&       non-MC &         8  \\
054&        1999/08/01 03:35&         1999/08/02 03:50&          24.3&       non-MC &         0  \\
055&        1999/08/02 16:32&         1999/08/03 17:10&          24.6&       non-MC &         0  \\
056&        1999/08/09 09:36&         1999/08/10 17:44&          32.1&       MC     &         92 \\
057&        1999/08/12 08:55&         1999/08/14 00:15&          39.3&       non-MC &         25 \\
058&        1999/08/21 11:50&         1999/08/23 11:30&          47.7&       non-MC &         49 \\
059&        1999/09/15 07:20&         1999/09/15 19:42&          12.4&       MC     &         24 \\
060&        1999/09/21 19:46&         1999/09/22 11:44&          16.0&       MC     &         75 \\
061&        1999/09/22 19:08&         1999/09/24 04:56&          33.8&       non-MC &         23 \\
062&        1999/10/21 03:51&         1999/10/22 05:56&          26.1&       non-MC &         35 \\
063&        1999/11/12 09:10&         1999/11/13 19:39&          34.5&       non-MC &         0  \\
064&        1999/11/14 00:42&         1999/11/14 23:26&          22.7&       non-MC &         0  \\
065&        1999/11/22 00:48&         1999/11/23 01:00&          24.2&       non-MC &         56 \\
066&        1999/11/23 06:12&         1999/11/23 18:42&          12.5&       non-MC &         0  \\
067&        1999/11/23 19:20&         1999/11/24 06:20&          11.0&       MC     &         0  \\

\hline
\end{tabular}
\end{flushleft}

\end{table}

\clearpage

\begin{table}

\renewcommand{\thetable}{1}

\caption{(Continued)}
\begin{flushleft}
\begin{tabular}{lcccccc}
\hline
\multicolumn{1}{c}{NO.}&Start$^{a}$&End$^{b}$&Duration$^{c}$&Type$^{d}$&Percent$^{e}$\\
\hline
068&        1999/12/12 19:32&         1999/12/13 16:28&          20.9&       non-MC &         98 \\
069&        1999/12/14 03:40&         1999/12/14 19:21&          15.7&       MC     &         74 \\
070&        1999/12/27 18:40&         1999/12/28 04:43&          10.1&       non-MC &         46 \\
071&        2000/01/22 17:00&         2000/01/25 08:20&          61.3&       MC     &         80 \\
072&        2000/02/12 12:15&         2000/02/13 00:10&          11.9&       MC     &         59 \\
073&        2000/02/14 12:18&         2000/02/16 07:51&          43.6&       non-MC &         0  \\
074&        2000/02/21 05:21&         2000/02/22 12:13&          30.9&       MC     &         74 \\
075&        2000/03/01 03:08&         2000/03/02 02:30&          23.4&       non-MC &         100\\
076&        2000/03/19 03:32&         2000/03/19 15:20&          11.8&       non-MC &         0  \\
077&        2000/03/28 03:10&         2000/03/29 19:28&          40.3&       non-MC &         0  \\
078&        2000/03/30 01:27&         2000/03/30 13:00&          11.6&       non-MC &         0  \\
079&        2000/03/31 02:58&         2000/04/01 02:15&          23.3&       non-MC &         6  \\
080&        2000/04/07 08:26&         2000/04/08 05:32&          21.1&       non-MC &         45 \\
081&        2000/04/18 20:05&         2000/04/19 13:45&          17.7&       non-MC &         0  \\
082&        2000/05/07 06:05&         2000/05/08 00:10&          18.1&       non-MC &         31 \\
083&        2000/05/13 16:51&         2000/05/14 17:48&          25.0&       non-MC &         0  \\
084&        2000/05/15 18:32&         2000/05/16 14:51&          20.3&       non-MC &         0  \\
085&        2000/05/23 08:58&         2000/05/23 20:32&          11.6&       non-MC &         0  \\
086&        2000/05/24 11:56&         2000/05/26 16:00&          52.1&       non-MC &         16 \\
087&        2000/06/04 23:50&         2000/06/06 22:00&          46.2&       non-MC &         38 \\
088&        2000/06/08 15:15&         2000/06/10 17:05&          49.8&       non-MC &         100\\
089&        2000/06/13 12:08&         2000/06/14 06:25&          18.3&       non-MC &         5  \\
090&        2000/06/24 06:32&         2000/06/26 00:12&          41.7&       non-MC &         0  \\
091&        2000/06/26 10:14&         2000/06/26 23:28&          13.2&       non-MC &         0  \\
092&        2000/07/01 07:30&         2000/07/03 08:43&          49.2&       MC     &         6  \\
093&        2000/07/11 22:48&         2000/07/13 02:16&          27.5&       non-MC &         35 \\
094&        2000/07/13 12:28&         2000/07/14 15:00&          26.5&       non-MC &         17 \\
095&        2000/07/14 17:17&         2000/07/15 14:15&          21.0&       non-MC &         0  \\
096&        2000/07/15 19:52&         2000/07/16 23:22&          27.5&       MC     &         47 \\
097&        2000/07/20 09:10&         2000/07/21 07:08&          22.0&       non-MC &         0  \\
098&        2000/07/23 19:20&         2000/07/26 02:42&          55.4&       non-MC &         0  \\
099&        2000/07/27 08:52&         2000/07/27 21:12&          12.3&       MC     &         8  \\
100&        2000/07/28 13:03&         2000/07/29 10:12&          21.2&       MC     &         38 \\
101&        2000/08/10 19:21&         2000/08/11 18:10&          22.8&       non-MC &         100\\
102&        2000/08/12 05:19&         2000/08/13 22:11&          40.9&       MC     &         82 \\
103&        2000/09/02 21:48&         2000/09/03 12:50&          15.0&       non-MC &         0  \\
104&        2000/09/06 00:44&         2000/09/06 16:12&          15.5&       non-MC &         0  \\
105&        2000/09/17 23:20&         2000/09/19 03:10&          27.8&       MC     &         100\\
106&        2000/10/03 15:08&         2000/10/05 02:33&          35.4&       MC     &         43 \\
107&        2000/10/05 16:36&         2000/10/07 06:28&          37.9&       non-MC &         17 \\
108&        2000/10/13 16:17&         2000/10/14 17:03&          24.8&       MC     &         73 \\
109&        2000/10/28 21:08&         2000/10/29 22:20&          25.2&       MC     &         100\\
110&        2000/11/06 22:15&         2000/11/07 17:22&          19.1&       MC     &         100\\
111&        2000/11/08 13:20&         2000/11/09 14:31&          25.2&       non-MC &         15 \\
112&        2000/11/11 08:11&         2000/11/12 00:08&          16.0&       non-MC &         5  \\
113&        2000/11/27 08:10&         2000/11/28 02:45&          18.6&       non-MC &         100\\
114&        2000/11/28 22:26&         2000/11/29 21:14&          22.8&       non-MC &         68 \\
115&        2000/12/03 13:10&         2000/12/05 07:41&          42.5&       non-MC &         5  \\
116&        2000/12/22 06:42&         2000/12/22 18:46&          12.1&       non-MC &         0  \\
117&        2000/12/23 00:48&         2000/12/23 11:53&          11.1&       non-MC &         14 \\
118&        2001/01/24 08:42&         2001/01/24 20:15&          11.6&       non-MC &         76 \\
119&        2001/03/04 05:08&         2001/03/05 01:38&          20.5&       non-MC &         0  \\
120&        2001/03/19 19:39&         2001/03/21 23:42&          52.1&       MC     &         25 \\
121&        2001/03/27 21:46&         2001/03/28 08:18&          10.5&       MC     &         52 \\
122&        2001/03/28 17:12&         2001/03/30 18:03&          48.9&       non-MC &         0  \\
123&        2001/03/31 05:31&         2001/03/31 21:39&          16.1&       non-MC &         0  \\
124&        2001/04/01 05:12&         2001/04/03 16:08&          58.9&       non-MC &         50 \\
125&        2001/04/04 18:02&         2001/04/05 10:31&          16.5&       MC     &         33 \\
126&        2001/04/08 13:16&         2001/04/09 03:31&          14.3&       non-MC &         24 \\
127&        2001/04/12 08:10&         2001/04/13 07:08&          23.0&       MC     &         100\\
128&        2001/04/13 10:31&         2001/04/14 11:08&          24.6&       non-MC &         48 \\
129&        2001/04/18 11:52&         2001/04/20 11:17&          47.4&       non-MC &         0  \\
130&        2001/04/21 23:38&         2001/04/23 03:02&          27.4&       MC     &         8  \\
131&        2001/04/28 16:38&         2001/05/01 21:43&          77.1&       MC     &         0  \\
132&        2001/05/07 17:50&         2001/05/08 07:48&          14.0&       MC     &         5  \\
133&        2001/05/09 11:54&         2001/05/10 21:19&          33.4&       non-MC &         0  \\
134&        2001/05/11 13:05&         2001/05/12 00:07&          11.0&       non-MC &         0  \\
135&        2001/05/28 04:38&         2001/05/29 21:26&          40.8&       MC     &         63 \\
136&        2001/05/29 21:26&         2001/05/31 14:52&          41.4&       non-MC &         0  \\

\hline
\end{tabular}
\end{flushleft}

\end{table}

\clearpage

\begin{table}

\renewcommand{\thetable}{1}

\caption{(Continued)}
\begin{flushleft}
\begin{tabular}{lcccccc}
\hline
\multicolumn{1}{c}{NO.}&Start$^{a}$&End$^{b}$&Duration$^{c}$&Type$^{d}$&Percent$^{e}$\\
\hline

137&        2001/06/07 17:19&         2001/06/08 06:52&          13.6&       non-MC &         0  \\
138&        2001/06/18 23:40&         2001/06/19 14:02&          14.4&       MC     &         0  \\
139&        2001/06/27 03:02&         2001/06/28 16:56&          37.9&       non-MC &         13 \\
140&        2001/07/09 02:22&         2001/07/11 04:18&          49.9&       MC     &         64 \\
141&        2001/08/15 06:00&         2001/08/16 14:31&          32.5&       non-MC &         0  \\
142&        2001/08/18 01:28&         2001/08/19 05:45&          28.3&       non-MC &         45 \\
143&        2001/08/30 17:15&         2001/08/31 09:34&          16.3&       MC     &         83 \\
144&        2001/09/01 13:57&         2001/09/02 18:02&          28.1&       MC     &         6  \\
145&        2001/09/13 17:36&         2001/09/14 20:58&          27.4&       non-MC &         35 \\
146&        2001/09/24 00:16&         2001/09/24 19:13&          19.0&       non-MC &         0  \\
147&        2001/09/25 05:40&         2001/09/25 20:03&          14.4&       non-MC &         0  \\
148&        2001/09/26 09:52&         2001/09/27 01:36&          15.7&       non-MC &         0  \\
149&        2001/09/29 12:10&         2001/09/30 18:45&          30.6&       non-MC &         75 \\
150&        2001/09/30 22:26&         2001/10/01 11:00&          12.6&       non-MC &         29 \\
151&        2001/10/02 04:04&         2001/10/03 16:27&          36.4&       non-MC &         18 \\
152&        2001/10/22 00:13&         2001/10/22 18:44&          18.5&       MC     &         90 \\
153&        2001/10/27 00:40&         2001/10/28 02:42&          26.0&       non-MC &         73 \\
154&        2001/10/29 21:32&         2001/10/31 12:51&          39.3&       non-MC &         22 \\
155&        2001/10/31 20:38&         2001/11/01 14:26&          17.8&       MC     &         100\\
156&        2001/11/14 05:12&         2001/11/15 14:01&          32.8&       non-MC &         0  \\
157&        2001/11/24 16:46&         2001/11/25 16:07&          23.4&       MC     &         84 \\
158&        2001/12/27 23:40&         2001/12/29 04:47&          29.1&       non-MC &         43 \\
159&        2001/12/30 02:23&         2001/12/30 19:31&          17.1&       MC     &         0  \\
160&        2002/02/28 16:49&         2002/03/01 09:43&          16.9&       non-MC &         30 \\
161&        2002/03/19 04:46&         2002/03/20 13:05&          32.3&       MC     &         0  \\
162&        2002/03/21 16:37&         2002/03/22 06:27&          13.8&       non-MC &         68 \\
163&        2002/03/22 13:45&         2002/03/23 10:53&          21.1&       non-MC &         0  \\
164&        2002/03/24 11:50&         2002/03/25 14:02&          26.2&       MC     &         57 \\
165&        2002/04/12 01:10&         2002/04/13 13:19&          36.2&       non-MC &         11 \\
166&        2002/04/17 23:08&         2002/04/19 08:02&          32.9&       MC     &         85 \\
167&        2002/04/20 04:42&         2002/04/21 15:23&          34.7&       non-MC &         66 \\
168&        2002/05/11 13:18&         2002/05/12 01:13&          11.9&       non-MC &         0  \\
169&        2002/05/19 02:45&         2002/05/20 02:55&          24.2&       MC     &         33 \\
170&        2002/05/20 09:08&         2002/05/21 20:54&          35.8&       MC     &         78 \\
171&        2002/05/23 21:36&         2002/05/25 17:48&          44.2&       MC     &         56 \\
172&        2002/07/18 12:00&         2002/07/19 09:31&          21.5&       MC     &         85 \\
173&        2002/07/20 02:28&         2002/07/22 04:53&          50.4&       non-MC &         0  \\
174&        2002/08/01 08:45&         2002/08/01 22:20&          13.6&       MC     &         33 \\
175&        2002/08/02 06:12&         2002/08/04 00:30&          42.3&       MC     &         35 \\
176&        2002/08/19 18:21&         2002/08/21 17:12&          46.9&       non-MC &         37 \\
177&        2002/09/07 16:08&         2002/09/08 18:38&          26.5&       non-MC &         43 \\
178&        2002/09/08 22:10&         2002/09/10 20:32&          46.4&       non-MC &         32 \\
179&        2002/09/19 21:00&         2002/09/20 22:15&          25.3&       non-MC &         46 \\
180&        2002/09/30 21:18&         2002/10/01 15:23&          18.1&       MC     &         0  \\
181&        2002/10/03 04:42&         2002/10/04 18:20&          37.6&       non-MC &         0  \\
182&        2002/11/17 16:30&         2002/11/19 12:48&          44.3&       MC     &         96 \\
183&        2002/12/17 20:14&         2002/12/19 01:31&          29.3&       non-MC &         93 \\
184&        2002/12/21 02:43&         2002/12/22 18:08&          39.4&       non-MC &         0  \\
185&        2003/01/27 01:58&         2003/01/27 23:38&          21.7&       non-MC &         65 \\
186&        2003/02/01 18:06&         2003/02/03 07:22&          37.3&       non-MC &         21 \\
187&        2003/02/18 04:08&         2003/02/19 15:45&          35.6&       non-MC &         0  \\
188&        2003/03/20 12:26&         2003/03/20 23:00&          10.6&       MC     &         35 \\
189&        2003/05/09 07:04&         2003/05/11 23:01&          64.0&       non-MC &         8  \\
190&        2003/05/29 18:32&         2003/05/30 15:53&          21.4&       non-MC &         22 \\
191&        2003/05/30 23:42&         2003/05/31 22:23&          22.7&       non-MC &         31 \\
192&        2003/06/15 21:38&         2003/06/16 20:31&          22.9&       non-MC &         28 \\
193&        2003/06/17 11:53&         2003/06/18 08:43&          20.8&       non-MC &         0  \\
194&        2003/07/06 12:25&         2003/07/07 11:42&          23.3&       non-MC &         0  \\
195&        2003/07/23 14:00&         2003/07/24 13:52&          23.9&       non-MC &         0  \\
196&        2003/08/04 23:51&         2003/08/06 01:08&          25.3&       MC     &         33 \\
197&        2003/08/16 02:00&         2003/08/17 13:40&          35.7&       non-MC &         8  \\
198&        2003/08/18 01:51&         2003/08/19 14:30&          36.7&       MC     &         78 \\
199&        2003/10/22 16:39&         2003/10/24 02:28&          33.8&       MC     &         100\\
200&        2003/10/24 18:32&         2003/10/25 11:24&          16.9&       non-MC &         66 \\
201&        2003/10/25 13:55&         2003/10/26 08:08&          18.2&       non-MC &         76 \\
202&        2003/10/26 18:32&         2003/10/28 01:30&          31.0&       non-MC &         28 \\
203&        2003/10/29 11:21&         2003/10/30 16:18&          29.0&       MC     &         31 \\
204&        2003/10/31 02:18&         2003/11/01 17:20&          39.0&       non-MC &         80 \\

\hline
\end{tabular}
\end{flushleft}

\end{table}

\clearpage

\begin{table}

\renewcommand{\thetable}{1}

\caption{(Continued)}
\begin{flushleft}
\begin{tabular}{lcccccc}
\hline
\multicolumn{1}{c}{NO.}&Start$^{a}$&End$^{b}$&Duration$^{c}$&Type$^{d}$&Percent$^{e}$\\
\hline
205&        2003/11/20 10:06&         2003/11/21 00:20&          14.2&       MC     &         0  \\
206&        2004/01/10 06:00&         2004/01/11 04:40&          22.7&       non-MC &         0  \\
207&        2004/01/22 09:47&         2004/01/23 14:16&          28.5&       MC     &         71 \\
208&        2004/01/24 07:33&         2004/01/25 02:57&          19.4&       MC     &         61 \\
209&        2004/02/17 18:14&         2004/02/18 15:51&          21.6&       non-MC &         0  \\
210&        2004/04/03 23:56&         2004/04/05 13:30&          37.6&       MC     &         51 \\
211&        2004/04/26 17:15&         2004/04/27 19:20&          26.1&       non-MC &         0  \\
212&        2004/05/01 00:27&         2004/05/01 11:49&          11.4&       non-MC &         0  \\
213&        2004/05/01 14:56&         2004/05/02 20:55&          30.0&       MC     &         77 \\
214&        2004/06/25 11:57&         2004/06/26 09:27&          21.5&       non-MC &         0  \\
215&        2004/07/22 20:08&         2004/07/24 05:37&          33.5&       MC     &         87 \\
216&        2004/07/24 16:50&         2004/07/25 12:10&          19.3&       MC     &         21 \\
217&        2004/07/25 21:10&         2004/07/26 22:27&          25.3&       MC     &         13 \\
218&        2004/07/27 02:12&         2004/07/27 18:34&          16.4&       MC     &         61 \\
219&        2004/08/01 08:32&         2004/08/02 04:28&          19.9&       non-MC &         0  \\
220&        2004/08/29 18:50&         2004/08/30 20:15&          25.4&       MC     &         54 \\
221&        2004/09/14 20:25&         2004/09/16 11:30&          39.1&       non-MC &         28 \\
222&        2004/09/18 12:18&         2004/09/19 23:57&          35.7&       non-MC &         57 \\
223&        2004/11/08 04:21&         2004/11/09 09:12&          28.9&       MC     &         71 \\
224&        2004/11/09 20:26&         2004/11/11 05:45&          33.3&       MC     &         9  \\
225&        2004/11/12 08:46&         2004/11/13 07:17&          22.5&       non-MC &         54 \\
226&        2004/12/12 22:31&         2004/12/13 19:16&          20.8&       MC     &         50 \\
227&        2004/12/27 16:26&         2004/12/29 04:16&          35.8&       non-MC &         83 \\
228&        2005/01/07 15:09&         2005/01/08 12:00&          20.9&       MC     &         28 \\
229&        2005/01/08 11:40&         2005/01/09 17:49&          30.2&       MC     &         29 \\
230&        2005/01/16 13:50&         2005/01/17 07:15&          17.4&       MC     &         17 \\
231&        2005/01/19 00:36&         2005/01/20 03:08&          26.5&       non-MC &         79 \\
232&        2005/02/18 15:08&         2005/02/19 05:15&          14.1&       non-MC &         98 \\
233&        2005/02/20 12:34&         2005/02/21 01:36&          13.0&       non-MC &         0  \\
234&        2005/02/21 04:08&         2005/02/22 02:43&          22.6&       non-MC &         0  \\
235&        2005/02/22 14:56&         2005/02/23 18:18&          27.4&       non-MC &         7  \\
236&        2005/05/15 05:30&         2005/05/17 09:32&          52.0&       MC     &         95 \\
237&        2005/05/17 09:32&         2005/05/19 02:51&          41.3&       non-MC &         61 \\
238&        2005/05/20 03:04&         2005/05/21 15:40&          36.6&       MC     &         11 \\
239&        2005/05/28 23:23&         2005/05/29 14:50&          15.5&       non-MC &         32 \\
240&        2005/05/30 01:05&         2005/05/30 15:19&          14.2&       non-MC &         67 \\
241&        2005/05/31 04:00&         2005/06/01 02:53&          22.9&       non-MC &         0  \\
242&        2005/06/12 15:00&         2005/06/13 11:42&          20.7&       non-MC &         0  \\
243&        2005/06/15 05:10&         2005/06/16 08:08&          27.0&       MC     &         56 \\
244&        2005/06/16 17:00&         2005/06/17 19:05&          26.1&       non-MC &         33 \\
245&        2005/07/10 10:18&         2005/07/12 04:23&          42.1&       non-MC &         45 \\
246&        2005/08/08 23:12&         2005/08/09 11:33&          12.4&       non-MC &         97 \\
247&        2005/08/24 20:33&         2005-08-25 13:08&          16.6&       non-MC &         74 \\
248&        2005/09/02 18:29&         2005/09/03 04:27&          10.0&       MC     &         74 \\
249&        2005/09/11 05:32&         2005/09/12 06:02&          24.5&       non-MC &         94 \\
250&        2005/09/12 20:39&         2005/09/13 13:31&          16.9&       non-MC &         57 \\
251&        2005/09/15 15:56&         2005/09/16 17:37&          25.7&       non-MC &         18 \\
252&        2005/09/20 20:50&         2005/09/21 17:56&          21.1&       non-MC &         0  \\
253&        2005/10/31 01:55&         2005/10/31 18:33&          16.6&       MC     &         33 \\
254&        2005/12/31 14:45&         2006/01/01 13:28&          22.7&       MC     &         100\\
255&        2006/02/05 19:16&         2006/02/06 11:35&          16.3&       MC     &         42 \\
256&        2006/04/13 15:45&         2006/04/14 11:02&          19.3&       MC     &         50 \\
257&        2006/07/10 20:06&         2006/07/11 17:42&          21.6&       non-MC &         30 \\
258&        2006/08/20 14:05&         2006/08/21 15:51&          25.8&       non-MC &         0  \\

\hline
\end{tabular}
\end{flushleft}

\end{table}

\clearpage
\begin{table}

\renewcommand{\thetable}{1}

\caption{(Continued)}
\begin{flushleft}
\begin{tabular}{lcccccc}
\hline
\multicolumn{1}{c}{NO.}&Start$^{a}$&End$^{b}$&Duration$^{c}$&Type$^{d}$&Percent$^{e}$\\
\hline

259&        2006/09/30 08:08&         2006/09/30 19:40&          11.5&       MC     &         36 \\
260&        2006/11/01 17:03&         2006/11/02 13:32&          20.5&       non-MC &         92 \\
261&        2006/11/18 11:02&         2006/11/20 02:10&          39.1&       non-MC &         0  \\
262&        2006/11/29 10:20&         2006/11/30 09:00&          22.7&       MC     &         18 \\
263&        2006/12/14 22:34&         2006/12/15 19:21&          20.8&       MC     &         90 \\
264&        2006/12/15 20:48&         2006/12/16 1900 &          22.2&       non-MC &         0  \\
265&        2006/12/17 01:35&         2006/12/17 23:49&          22.2&       non-MC &         0  \\
266&        2007/01/14 11:47&         2007/01/15 07:06&          19.3&       MC     &         15 \\
267&        2007/05/21 22:22&         2007/05/22 13:12&          14.8&       MC     &         59 \\
268&        2007/11/19 23:12&         2007/11/20 11:39&          12.5&       MC     &         0  \\
269&        2008/09/17 03:52&         2008/09/18 07:47&          27.9&       MC     &         88 \\
270&        2008/10/08 04:30&         2008/10/08 19:43&          15.2&       non-MC &         0  \\
271&        2008/12/04 12:37&         2008/12/05 11:00&          22.4&       non-MC &         0  \\
272&        2008/12/17 03:38&         2008/12/17 17:28&          13.8&       MC     &         0  \\
\hline
\end{tabular}
\end{flushleft}
a {The beginning of the ICME (UT).}

b {The end of the ICME (UT).}

c {The duration of the ICME (h).}

d {The type of the ICME.}

e {The percentages of counterstreaming intervals coincident with the ICME (\%).}

\end{table}


\clearpage

\begin{thebibliography}{}
\bibitem[Awasthi et al.(2018)]{2018ApJ...857..124A} Awasthi, A.~K., Liu, R., Wang, H., et al.\ 2018, \apj, 857, 124
\bibitem[Burlaga et al.(1981)]{1981JGR....86.6673B} Burlaga, L., Sittler, E., Mariani, F., \& Schwenn, R.\ 1981, \jgr, 86, 6673
\bibitem[Burlaga et al.(2002)]{2002JGRA..107.1266B} Burlaga, L.~F., Plunkett, S.~P., \& St. Cyr, O.~C.\ 2002, Journal of Geophysical Research (Space Physics), 107, 1266
\bibitem[Burlaga et al.(2001)]{2001JGR...10620957B} Burlaga, L.~F., Skoug, R.~M., Smith, C.~W., et al.\ 2001, \jgr, 106, 20957
\bibitem[Cane et al.(1997)]{1997JGR...102.7075C} Cane, H.~V., Richardson, I.~G., \& Wibberenz, G.\ 1997, \jgr, 102, 7075
\bibitem[Cane, \& Richardson(2003)]{2003JGRA..108.1156C} Cane, H.~V., \& Richardson, I.~G.\ 2003, Journal of Geophysical Research (Space Physics), 108, 1156
\bibitem[Canfield et al.(1999)]{1999GeoRL..26..627C} Canfield, R.~C., Hudson, H.~S., \& McKenzie, D.~E.\ 1999, \grl, 26, 627
\bibitem[Chi et al.(2016)]{2016SoPh..291.2419C} Chi, Y., Shen, C., Wang, Y., et al.\ 2016, \solphys, 291, 2419
\bibitem[Crooker et al.(2004)]{2004JGRA..109.6110C} Crooker, N.~U., Forsyth, R., Rees, A., et al.\ 2004, Journal of Geophysical Research (Space Physics), 109, A06110
\bibitem[Feng et al.(2006)]{2006JGRA..111.7S90F} Feng, H.~Q., Wu, D.~J., \& Chao, J.~K.\ 2006, Journal of Geophysical Research (Space Physics), 111, A07S90
\bibitem[Feng et al.(2007)]{2007JGRA..112.2102F} Feng, H.~Q., Wu, D.~J., \& Chao, J.~K.\ 2007, Journal of Geophysical Research (Space Physics), 112, A02102
\bibitem[Feng et al.(2008)]{2008JGRA..11312105F} Feng, H.~Q., Wu, D.~J., Lin, C.~C., et al.\ 2008, Journal of Geophysical Research (Space Physics), 113, A12105
\bibitem[Feng et al.(2010)]{2010JGRA..11510109F} Feng, H.~Q., Wu, D.~J., \& Chao, J.~K.\ 2010, Journal of Geophysical Research (Space Physics), 115, A10109
\bibitem[Feng et al.(2015)]{2015JGRA..12010175F} Feng, H.~Q., Zhao, G.~Q., \& Wang, J.~M.\ 2015, Journal of Geophysical Research (Space Physics), 120, 10
\bibitem[Feng \& Wang(2015)]{2015ApJ...809..112F} Feng, H.~Q., \& Wang, J.~M.\ 2015, \apj, 809, 112
\bibitem[Feng et al.(2019)]{2019SCTS190310938F} Feng, H.~Q., Zhao, G.~Q., \& Wang, J.~M.\ 2019,  Sci. China Technological Sci., https://doi.org/10.1007/s11431-018-9481-1.
\bibitem[Feldman et al.(1975)]{1975JGR....80.4181F} Feldman, W.~C., Asbridge, J.~R., Bame, S.~J., et al.\ 1975, \jgr, 80, 4181
\bibitem[Feldman et al.(1982)]{1982JGR....87..632F} Feldman, W.~C., Anderson, R.~C., Asbridge, J.~R., et al.\ 1982, \jgr, 87, 632
\bibitem[Gopalswamy et al.(2001)]{2001ApJ...548L..91G} Gopalswamy, N., Yashiro, S., Kaiser, M.~L., et al.\ 2001, \apjl, 548, L91
\bibitem[Gopalswamy et al.(2013a)]{2013aSoPh..284....1G} Gopalswamy, N., Nieves-Chinchilla, T., Hidalgo, M., et al.\ 2013a, \solphys, 284, 1
\bibitem[Gopalswamy et al.(2013b)]{2013bSoPh..284...17G} Gopalswamy, N., M{\"a}kel{\"a}, P., Akiyama, S., et al.\ 2013b, \solphys, 284, 17
\bibitem[Gosling et al.(1973)]{1973JGR....78.2001G} Gosling, J.~T., Pizzo, V., \& Bame, S.~J.\ 1973, \jgr, 78, 2001
\bibitem[Gosling et al.(1987)]{1987JGR....92.8519G} Gosling, J.~T., Baker, D.~N., Bame, S.~J., et al.\ 1987, \jgr, 92, 8519
\bibitem[Gosling(1990)]{1990GMS....58..343G} Gosling, J.~T.\ 1990, Washington DC American Geophysical Union Geophysical Monograph Series, 58, 343
\bibitem[Gosling et al.(1993)]{1993GeoRL..20.2335G} Gosling, J.~T., Bame, S.~J., Feldman, W.~C., et al.\ 1993, \grl, 20, 2335
\bibitem[Gosling et al.(1995)]{1995GeoRL..22..869G} Gosling, J.~T., Birn, J., \& Hesse, M.\ 1995, \grl, 22, 869
\bibitem[Gosling et al.(2001)]{2001GeoRL..28.4155G} Gosling, J.~T., Skoug, R.~M., \& Feldman, W.~C.\ 2001, \grl, 28, 4155
\bibitem[Gosling et al.(2002)]{2002GeoRL..29.1573G} Gosling, J.~T., Skoug, R.~M., Feldman, W.~C., et al.\ 2002, \grl, 29, 1573
\bibitem[Hirshberg et al.(1972)]{1972SoPh...23..467H} Hirshberg, J., Bame, S.~J., \& Robbins, D.~E.\ 1972, \solphys, 23, 467
\bibitem[Hundhausen et al.(1984)]{1984JGR....89.2639H} Hundhausen, A.~J., Sawyer, C.~B., House, L., Illing, R.~M.~E., \& Wagner, W.~J.\ 1984, \jgr, 89, 2639 
\bibitem[Huttunen et al.(2002)]{2002JGRA..107.1121H} Huttunen, K.~E.~J., Koskinen, H.~E.~J., \& Schwenn, R.\ 2002, Journal of Geophysical Research (Space Physics), 107, 1121
\bibitem[Janvier et al.(2014)]{2014JGRA..119.7088J} Janvier, M., D{\'e}moulin, P., \& Dasso, S.\ 2014, Journal of Geophysical Research (Space Physics), 119, 7088
\bibitem[Kilpua et al.(2011)]{2011JASTP..73.1228K} Kilpua, E.~K.~J., Jian, L.~K., Li, Y., Luhmann, J.~G., \& Russell, C.~T.\ 2011, Journal of Atmospheric and Solar-Terrestrial Physics, 73, 1228
\bibitem[Kilpua et al.(2017)]{2017LRSP...14....5K} Kilpua, E., Koskinen, H.~E.~J., \& Pulkkinen, T.~I.\ 2017, Living Reviews in Solar Physics, 14, 5
\bibitem[Kim et al.(2013)]{2013SoPh..284...77K} Kim, R.-S., Gopalswamy, N., Cho, K.-S., et al.\ 2013, \solphys, 284, 77
\bibitem[Larson et al.(1997)]{1997GeoRL..24.1911L} Larson, D.~E., Lin, R.~P., McTiernan, J.~M., et al.\ 1997, \grl, 24, 1911
\bibitem[Lavraud et al.(2010)]{2010AnGeo..28..233L} Lavraud, B., Opitz, A., Gosling, J.~T., et al.\ 2010, Annales Geophysicae, 28, 233
\bibitem[Lepping et al.(1990)]{1990JGR....9511957L} Lepping, R.~P., Burlaga, L.~F., \& Jones, J.~A.\ 1990, \jgr, 95, 11957
\bibitem[Lepri et al.(2001)]{2001JGR...10629231L} Lepri, S.~T., Zurbuchen, T.~H., Fisk, L.~A., et al.\ 2001, \jgr, 106, 29231
\bibitem[Lepri, \& Zurbuchen(2004)]{2004JGRA..109.1112L} Lepri, S.~T., \& Zurbuchen, T.~H.\ 2004, Journal of Geophysical Research (Space Physics), 109, A01112
\bibitem[Liu et al.(2010)]{2010ApJ...725L..84L} Liu, R., Liu, C., Wang, S., et al.\ 2010, \apjl, 725, L84
\bibitem[Liu et al.(2014)]{2014ApJ...793L..41L} Liu, Y.~D., Yang, Z., Wang, R., et al.\ 2014, \apjl, 793, L41
\bibitem[Lynch et al.(2008)]{2008ApJ...683.1192L} Lynch, B.~J., Antiochos, S.~K., DeVore, C.~R., et al.\ 2008, \apj, 683, 1192
\bibitem[M{\"a}kel{\"a} et al.(2013)]{2013SoPh..284...59M} M{\"a}kel{\"a}, P., Gopalswamy, N., Xie, H., et al.\ 2013, \solphys, 284, 59
\bibitem[Manchester et al.(2017)]{2017SSRv..212.1159M} Manchester, W., Kilpua, E.~K.~J., Liu, Y.~D., et al.\ 2017, \ssr, 212, 1159
\bibitem[McComas et al.(1998)]{1998SSRv...86..563M} McComas, D.~J., Bame, S.~J., Barker, P., et al.\ 1998, \ssr, 86, 563
\bibitem[Pagel et al.(2005)]{2005JGRA..110.1103P} Pagel, C., Crooker, N.~U., Larson, D.~E., et al.\ 2005, Journal of Geophysical Research (Space Physics), 110, A01103
\bibitem[Reinard(2005)]{2005ApJ...620..501R} Reinard, A.\ 2005, \apj, 620, 501
\bibitem[Richardson \& Cane(2004)]{2004JGRA..109.9104R} Richardson, I.~G., \& Cane, H.~V.\ 2004, Journal of Geophysical Research (Space Physics), 109, A09104
\bibitem[Riley, \& Crooker(2004)]{2004ApJ...600.1035R} Riley, P., \& Crooker, N.~U.\ 2004, \apj, 600, 1035
\bibitem[Rodkin et al.(2018)]{2018SoPh..293...78R} Rodkin, D., Slemzin, V., Zhukov, A.~N., et al.\ 2018, \solphys, 293, 78
\bibitem[Rouillard et al.(2011)]{2011ApJ...734....7R} Rouillard, A.~P., Sheeley, N.~R., Jr., Cooper, T.~J., et al.\ 2011, \apj, 734, 7
\bibitem[Rosenbauer et al.(1977)]{1977JGZG...42..561R} Rosenbauer, H., Schwenn, R., Marsch, E., et al.\ 1977, Journal of Geophysics Zeitschrift Geophysik, 42, 561
\bibitem[Rust, \& Kumar(1996)]{1996ApJ...464L.199R} Rust, D.~M., \& Kumar, A.\ 1996, \apjl, 464, L199
\bibitem[Shodhan et al.(2000)]{2000JGR...10527261S} Shodhan, S., Crooker, N.~U., Kahler, S.~W., et al.\ 2000, \jgr, 105, 27261
\bibitem[Skoug et al.(2006)]{2006JGRA..111.1101S} Skoug, R.~M., Gosling, J.~T., McComas, D.~J., et al.\ 2006, Journal of Geophysical Research (Space Physics), 111, A01101
\bibitem[Stansberry et al.(1988)]{1988JGR....93.1975S} Stansberry, J.~A., Gosling, J.~T., Thomsen, M.~F., et al.\ 1988, \jgr, 93, 1975
\bibitem[Steinberg et al.(2005)]{2005JGRA..110.6103S} Steinberg, J.~T., Gosling, J.~T., Skoug, R.~M., et al.\ 2005, Journal of Geophysical Research (Space Physics), 110, A06103
\bibitem[Wang \& Feng(2016)]{2016SCTS591051W} Wang, J.~M. \&Feng, H.~Q.,\ 2016, Sci. China Earth Sci., 59, 1051
\bibitem[Wang et al.(2019)]{2019ApJ...876...57W} Wang, J.~M., Feng, H.~Q., Li, H.~B., et al.\ 2019, \apj, 876, 57
\bibitem[Webb et al.(2000)]{2000JGR...105.7491W} Webb, D.~F., Cliver, E.~W., Crooker, N.~U., et al.\ 2000, \jgr, 105, 7491
\bibitem[Yashiro et al.(2013)]{2013SoPh..284....5Y} Yashiro, S., Gopalswamy, N., M{\"a}kel{\"a}, P., et al.\ 2013, \solphys, 284, 5
\bibitem[Zhang et al.(2013)]{2013SoPh..284...89Z} Zhang, J., Hess, P., \& Poomvises, W.\ 2013, \solphys, 284, 89
\bibitem[Zhang et al.(2007)]{2007JGRA..11210102Z} Zhang, J., Richardson, I.~G., Webb, D.~F., et al.\ 2007, Journal of Geophysical Research (Space Physics), 112, A10102
\bibitem[Zwickl et al.(1982)]{1982JGR....87.7379Z} Zwickl, R.~D., Asbridge, J.~R., Bame, S.~J., et al.\ 1982, \jgr, 87, 7379
\end{thebibliography}
\end{document}